\documentclass{article}
\usepackage{graphicx}

\newcommand{\ket}[1]{|#1\rangle}

\begin{document}
 \title{{\bf Phase space methods for particles on a circle}}

 \author{S. Zhang and A. Vourdas\\
 Department of Computing,\\
 University of Bradford, \\
 Bradford BD7 1DP, United Kingdom}

 \maketitle

 \begin{abstract}
 The phase space $S\times Z$ for a particle on a circle is considered. Displacement
 operators in this phase space are introduced and their properties are studied.
 Wigner and Weyl functions in this context are also considered and their physical 
 interpretation and properties are discussed. All results are compared and contrasted
 with the corresponding ones for the harmonic oscillator in the $R \times R$ phase space.

 \end{abstract}

 \section{Introduction}
Since the work of Wigner \cite{Wig} and Moyal \cite{Moy}, phase
space methods have been used extensively in quantum mechanics. A
lot of this work is for the harmonic oscillator where both the
position and momentum take values in the real line $R$ and the
phase space is the plane $R\times R$. There has also been work on
finite quantum systems \cite{Wey,Fin}, where both the position and
momentum take values in $Z_N$ (the integers module N) and the
phase space is the lattice $Z_N \times Z_N$. The purpose of this
paper is to study phase space methods for quantum particles on a
circle. In this case the position takes values on a circle $S$ and
the momentum take discrete values in $Z$ (the integers times a
factor). In this case the phase space is $S \times Z$. We note
that in any area where there is Fourier transform involved (eg in
signal processing), the phase space can be $R \times R$ or $Z_N
\times Z_N$ or $S\times Z$ in the sense that where one of the
variables takes values in $R$ or $Z_N$ or $S$, the `dual variable'
take values in $R$ or $Z_N$ or $Z$, correspondingly.

Quantum mechanics on a circle is the simplest example of quantum mechanics
in a non-trivial topology and has been studied extensively in the literature \cite{cir1,cir2}.
Physical applications include Aharonov-Bohm phenomena \cite{ab}, mesoscopic Aharonov-
Bohm rings, Floquet-Bloch wavefunctions in solid state systems, etc.

In section 2 we introduce the basic formulism for position and
momentum states and operators, taking into account the non-trivial
topology of our system (described by the winding number). In
section 3 we introduce displacement operators and study their
properties. In section 4 we define Wigner and Weyl functions. We
show that the properties of the displacement operators lead to
analogous properties for the Wigner and Weyl functions. In section
5 we discuss an example based on a Theta wavefunction (which is
the analogue in a circle, of a Gaussian wavefunction in a real
line). Numerical examples for the corresponding Wigner and Weyl
functions are discussed. We conclude in section 6 with a
discussion of our results.

 \section{Position and momentum states}
 An electric charge is moving on a circle parameterized by the variable $x$.
 The winding number $w_x$ of $x$ is defined as the integer part of the $x/(2\pi r)$ (for negative $x$
 it is the integer part of the $x/(2\pi r)$ minus $1$).
 Let $r$ be the radius of the circle. A magnetostatic flux $\phi$ is threading the circle
 in the perpendicular direction.The wavefunction $R(x)$ obeys the quasi-periodic boundary condition
 (in units $K_B=\hbar=c=1$)
  \begin{equation}\label{Bloch}
   R(x+2\pi r)=R(x) \exp (ie\phi),
  \end{equation}
Similar functions also appear in solid state Physics
 (Bloch functions).
 They are normalizable within each period
  \begin{equation}\label{Norm}
   \frac {1}{2\pi r}\int_0^{2\pi r} |R(x)|^2 \textrm{d}x =1,
  \end{equation}

 Since $R(x)$ is a quasi-periodic function, it can be written as the following Fourier expansion:
  \begin{equation}\label{Rx}
   R(x)=\sum_{N=-\infty}^{\infty} R_N \exp (i p_N x);\;\;\;\;\;p_N=\frac {N+\sigma}{r};\;\;\;\;\;
   \sigma=\frac{e\phi}{2 \pi}.
  \end{equation}
 The inverse Fourier transform gives:
  \begin{equation}\label{Rn}
   R_N=\frac{1}{2\pi r} \int _0^{2\pi r} \exp (-ip_N x)R(x) \textrm{d}x.
  \end{equation}
 $R(x)$ and $R_N$ can be respectively considered as the position and momentum representations of the state
 $\ket{R}$. So Eq(\ref{Rx}) and Eq(\ref{Rn}) can be written as:
  \begin{equation}
   \langle x|R \rangle = \sum_{N=-\infty}^{\infty} \langle p_N|R \rangle \exp (i p_N x);
  \end{equation}
  \begin{equation}
   \langle p_N|R \rangle = \frac{1}{2\pi r} \int _0^{2\pi r} \langle x|R \rangle \exp (-ip_N x) \textrm{d}x.
  \end{equation}

 Let  $|x \rangle$, $|p_N \rangle$ be position and momentum eigenstates, correspondingly. Then:
  \begin{eqnarray}
   |x \rangle =\sum_{N=-\infty}^{\infty}\exp (-ip_N x)|p_N \rangle;\;\;\;\;
   |p_N \rangle =\frac{1}{2\pi r}\int _0 ^{2\pi r}dx\exp (ip_N x)|x \rangle  \label{x_pn}\\
   \langle x|y \rangle = (2\pi r) \delta \left [x-y+2\pi r(w_y-w_x) \right ] \exp[-i2\pi\sigma (w_y-w_x)];\;\;\;\;
   \langle p_M|p_N \rangle = \delta_{MN}.
  \end{eqnarray}
 It is easily seen that
  \begin{equation}
   |x+2\pi rw \rangle =\exp (-i2\pi \sigma w)|x \rangle,
  \end{equation}

 The completeness can can be expressed as:
  \begin{equation}
   \frac{1}{2\pi r} \int _0^{2\pi r} |x\rangle \langle x| \textrm{d}x =
   \sum_{N=-\infty}^{\infty} |p_N\rangle \langle p_N| = {\bf 1}.
  \end{equation}

Position and momentum operators are defined as:
  \begin{equation}
   \hat{x}=\frac{1}{2\pi r} \int _{0}^{2\pi r} x|x\rangle \langle x| \textrm{d}x,
   \;\;\;\;\;\;\;\;
   \hat{p}=\sum_{N=-\infty}^{\infty} p_N |p_N\rangle \langle p_N|.
  \end{equation}
 We note that a different definition of $\hat{x}$ that involves integration from $\tau$
 to $\tau+2\pi r$ leads to:
 \begin{equation}
  \hat{x}_{\tau}=\hat{x}+\Pi _{\tau};\;\;\;\;\;\Pi _{\tau}=\int _0^{\tau} |x\rangle \langle x| \textrm{d}x
  \end{equation}
 The $\hat{x}_{\tau}$ is different from $\hat{x}$ by the projection operator $\Pi _{\tau}$.
 In the special case that $\tau=2\pi rw$ where $w$ is an integer,  $\Pi _{2\pi r w}=2\pi rw{\bf 1}$.
 It is easily seen that:
  \begin{equation}\label{eigxpn}
   \hat{x}\ket{x}=(x-2\pi rw_x)|x\rangle, \;\;\;\;\;\;\;\; \hat{p} |p_N\rangle = p_N |p_N\rangle.
  \end{equation}

 \section{Displacements and parity}
 We define  displacement operators as
  \begin{eqnarray}
   D(\alpha,K)&\equiv & \exp \left (-\frac{i\alpha K}{2r} \right ) \exp \left (i\frac{K}{r}\hat{x} \right ) 
                       \exp \left (-i\alpha\hat{p} \right ), \\
   D(\alpha, K)\ket{x}&=& \exp \left [\frac{iK}{r} \left (x+\frac{\alpha}{2} \right ) \right ]|x+\alpha \rangle, \label{Dx}\\
   D(\alpha, K)\ket{p_N}&=&\exp \left (-\frac{i\alpha K}{2r} \right ) \exp(-i\alpha p_N)|p_{N+K} \rangle.  \label{Dp}
  \end{eqnarray}
 It is easily seen that
  \begin{equation}
   D(\alpha, K)D(\beta, M)=D(\alpha+\beta, K+M) \exp \left( \frac{K\beta}{2r}-\frac{M\alpha}{2r} \right)
  \end{equation}
  \begin{equation}\label{AQpD}
   D(\alpha + 2\pi rw, K) = (-1)^{Kw}\exp(-i2\pi \sigma w)D(\alpha,K),
  \end{equation}
  \begin{equation}
   D^\dagger(\alpha,K)=D(-\alpha,-K),
  \end{equation}
where $w$ is an integer (the winding number).
For later purposes we note that the
$D(\alpha,K) \exp \left (\frac{i\alpha\sigma}{r} \right )$ is periodic in $\alpha$. The period
is $2\pi r$ if $K$ is even and $4\pi r$ if $K$ is odd number:
\begin{equation}\label{QpD}
   D(\alpha + 2\pi r, K)\exp \left [ \frac{i(\alpha+2\pi r)\sigma}{r}\right ] = (-1)^{K} 
   \exp \left (\frac{i\alpha\sigma}{r} \right )D(\alpha,K),
  \end{equation}

 We also define the parity operator as:
 \begin{eqnarray}
    U_0 = \frac{1}{2\pi r} \int^{2\pi r}_{0} |\alpha\rangle \langle -\alpha| \exp \left (\frac{i2\alpha\sigma}{r} \right ) 
          \textrm{d}\alpha = \sum_{N=-\infty}^{\infty} |p_{-N}\rangle \langle p_N|.  \label{U0}
   \end{eqnarray}
The `flux factor' $\exp \left (\frac{i\alpha\sigma}{r} \right )$ has been included in the definition
so that the integrand $|\alpha\rangle \langle -\alpha| \exp \left (\frac{i2\alpha\sigma}{r} \right )$
is periodic.
The parity operator obeys the relations:
   \begin{equation}
    U_0 = U^\dagger_0, \;\;\;\;\;\;\;\; U^2_0=I;
   \end{equation}
The flux breaks the parity symmetry.
The parity operator acting on the state $|p_{-N}\rangle$ (which has momentum $\frac{-N+\sigma}{r}$ )
gives the state $|p_{N}\rangle$ (which has momentum $\frac{N+\sigma}{r}$ );
 i.e., it is a parity with momentum origin $\sigma/r$.
The parity operator acting on the state $|x\rangle$ gives the state $|-x\rangle \exp \left (-\frac{2ix\sigma}{r} \right )$.
The `flux factor' $\exp \left (-\frac{2ix\sigma}{r} \right )$ `corrects' the quasi-periodicity caused by the flux 
into periodicity.

We note here that in the harmonic oscillator case the displacement operators $D(z)= \exp(za^\dagger - z^* a) $
obey the important relations \cite{bv}
   \begin{eqnarray}
    \int_{-\infty}^{\infty} \textrm{d}z_R D(z) &=& \sqrt{2} \pi |p= \frac{z_I}{\sqrt{2}} \rangle 
                                                   \langle p=-\frac{z_I}{\sqrt{2}}|,  \label{HD1}\\
    \int_{-\infty}^{\infty} \textrm{d}z_I D(z) &=& \sqrt{2} \pi |x= \frac{z_R}{\sqrt{2}} \rangle 
                                                   \langle x=-\frac{z_R}{\sqrt{2}}|. \label{HD2}\\
    \int \frac{\textrm{d}^2z}{2\pi} D(z) &=& U_0. \label{HD3}
   \end{eqnarray}
   where $z=z_R+iz_I$  and $U_0$ is here the harmonic oscillator parity operator (defined in \cite{bv}).
   These relations are intimately related with the marginal properties of the Wigner and Weyl functions.

Motivated by this we study here similar relations in our context
of quantum mechanics on a circle. 
We first define the function
   \begin{eqnarray}
    \Delta (x)=\frac{1}{2\pi} \int_{0}^{2\pi} \exp(i\beta x) \textrm{d}\beta =\exp(-i\pi x)\frac{\sin(\pi x)}{\pi x}\equiv
    \exp(-i\pi x)\textrm{sinc}(x).
   \end{eqnarray}
This is the sinc-function (used extensively in areas like digital signal processing) with a phase factor.
   For integers $M,N$
   \begin{equation}
    \Delta(M-N) = \delta (M,N)
   \end{equation}
   where $\delta $ is Kronecker's delta.
Using Eqs(\ref{Dx}),(\ref{Dp})
can prove that:
   \begin{equation}\label{DdK}
    \sum_{K=-\infty}^{\infty} D(\alpha,K) = |\frac{\alpha}{2}\rangle \langle -\frac{\alpha}{2}|,
   \end{equation}
   \begin{equation}\label{Dda}
     \frac{1}{2\pi r} \int_{0}^{2\pi r} D(\alpha,K) \exp \left (\frac{i\alpha\sigma}{r} \right )
          \textrm{d}\alpha =
          \left\{  \begin{array}{ll}
                  |p_{-M}\rangle \langle p_M|, & K=2M; \\
                  \sum_{N=-\infty}^{\infty} |p_{N+K}\rangle \langle p_{N}| \Delta \left (-\frac{K}{2}-N \right ), & K=2M+1.
                \end{array}
          \right.
   \end{equation}
We have explained earlier why the integrations  involve the
`flux factor' $\exp \left (\frac{i\alpha\sigma}{r} \right )$.
It is natural to ask what is the result in Eq(\ref{Dda}) if we do not include the `flux factor'.
We can prove 
  \begin{equation}
   \frac{1}{2\pi r} \int_0^{2\pi r} D(\alpha,K) \textrm{d}\alpha
   = \sum_{N=-\infty}^{\infty} |p_{N+K}\rangle \langle p_{N}| \Delta \left (-\frac{K}{2}-N-\sigma \right ).
  \end{equation}
In Eq(\ref{Dda}), integration from $2\pi r$ to $4\pi r$, gives the same result for even $K$
and the same result but with opposite sign for odd $K$:
 \begin{equation}\label{DdaK}
     \frac{1}{2\pi r} \int_{2\pi r}^{4\pi r} D(\alpha,K) \exp \left (\frac{i\alpha\sigma}{r} \right )
          \textrm{d}\alpha =
          \left\{  \begin{array}{ll}
                  |p_{-M}\rangle \langle p_M|, & K=2M; \\
                  -\sum_{N=-\infty}^{\infty} |p_{N+K}\rangle \langle p_{N}| \Delta \left (-\frac{K}{2}-N \right ), & K=2M+1.
                \end{array}
          \right.
   \end{equation}

This is related to the fact (Eq(\ref{QpD})) that for odd $K$ the $D(\alpha,K) \exp \left (\frac{i\alpha\sigma}{r} \right )$
is `anti-periodic' with `period' $2\pi r$ (it is periodic with period $4\pi r$).
 Therefore if we integrate from $0$ to $4\pi r$
we get the same result for even $K$ and zero for odd $K$.

Combining Eqs(\ref{U0}),(\ref{Dda}) we easily show that:
 \begin{equation} \label{DdaKev}
     \sum_{K=even} \frac{1}{2\pi r} \int^{2\pi r}_{0} D(\alpha,K)
           \exp \left (\frac{i\alpha\sigma}{r} \right ) \textrm{d}\alpha =U_0
   \end{equation}
It is natural to ask what is the result in Eq(\ref{DdaKev}) if we sum over both even and odd integers. 
We can prove 
  \begin{equation}
    \sum_{K=-\infty}^{\infty} \frac{1}{2\pi r} \int_0^{2\pi r}  D(\alpha,K)
    \exp \left (\frac{i\alpha\sigma}{r} \right ) \textrm{d}\alpha = 
     U_0 + \sum_{M=-\infty}^{\infty} \sum_{N=-\infty}^{\infty}
    |p_{N+2M-1}\rangle \langle p_{N}| \Delta \left (-M-N+\frac{1}{2} \right );
  \end{equation}

The displaced parity operator is
   \begin{equation}\label{UaK}
    U(\alpha,K) = D(\alpha,K) U_0 D^\dagger (\alpha,K) =\exp \left (\frac{i\alpha\sigma}{r} \right ) D(2\alpha,2K) U_0 \\
                = \exp \left (-\frac{i\alpha\sigma}{r} \right ) U_0 D(-2\alpha,-2K).
   \end{equation}
  $U(\alpha,K)$ is a periodic function of $\alpha$ with the period $\pi r$:
   \begin{equation}\label{PU}
    U(\alpha+\pi r,K)=U(\alpha,K).
   \end{equation}
 We next show that:
   \begin{equation}\label{UdK}
    \sum_{K=-\infty}^{\infty} U(\alpha,K) = \frac{1}{2} \left (|\alpha\rangle \langle \alpha|
                      + |\alpha+\pi r\rangle \langle \alpha+\pi r| \right ),
   \end{equation}
   \begin{equation}\label{Uda}
     \frac{1}{2\pi r} \int^{2\pi r}_{0} U(\alpha,K) \textrm{d}\alpha =
                            |p_K\rangle \langle p_K|
   \end{equation}
   \begin{equation}\label{UI}
     \sum_{K} \frac{1}{2\pi r} \int^{2\pi r}_{0} U(\alpha,K)
           \textrm{d}\alpha ={\bf 1}
   \end{equation}
 Eq(\ref{UdK}) is proved if we use Eq(\ref{DdK}),(\ref{UaK}) in conjunction with the fact that
 \begin{equation}\label{evenD}
   \sum_{K= \textrm{even}} D(\alpha,K) = \frac{1}{2} \left (|\frac{\alpha}{2}\rangle \langle -\frac{\alpha}{2}| +
                                          |\frac{\alpha}{2}+ \pi r\rangle \langle -\frac{\alpha}{2} + \pi r| \right ).
 \end{equation}
For completeness we also mention that 
 \begin{equation}\label{evenD1}
   \sum_{K= \textrm{odd}} D(\alpha,K) = \frac{1}{2} \left (|\frac{\alpha}{2}\rangle \langle -\frac{\alpha}{2}| -
                                          |\frac{\alpha}{2}+ \pi r\rangle \langle -\frac{\alpha}{2} + \pi r| \right ).
 \end{equation}
Eq(\ref{evenD1}) can be proved using Eq(\ref{Dx}).
We note that in Eq(\ref{UdK}) we have projectors to the diametrically opposite position states $ |\alpha\rangle $
and  $ |\alpha +\pi r\rangle $. This is related to the fact that the displaced parity operator
on the left hand side of Eq(\ref{UdK}), has
 period $\pi r$. Eq(\ref{Uda}) can be proved using Eq(\ref{Dda}). Summation over $K$ in Eq(\ref{Uda}) gives Eq(\ref{UI}).
 
 The displacement operators are related to the displaced parity operators through a two-dimensional Fourier transform.
 Multiplying the left and right hand sides of Eq(\ref{DdaKev}) with $D(\alpha, K)$ and $[D(\alpha, K)]^\dagger$ correspondingly,
 we can prove:
   \begin{equation} \label{QW}
     \sum_{M=even} \frac{1}{2\pi r} \int^{2\pi r}_{0} D(\beta,M)
       \exp \left (\frac{i\beta\sigma}{r} \right ) \exp \left [ i\frac{K\beta-M\alpha}{r} \right ] \textrm{d}\beta =U(\alpha,K)
   \end{equation}

 \section{Wigner and Weyl functions}

 The Weyl and Wigner functions can be defined in terms of the displacement and parity operator
 correspondingly, as:
  \begin{eqnarray}\label{Weyl}
   \tilde{W}(\alpha,K) &=& \textrm{Tr} \left [\hat{\rho}D(\alpha,K) \right ], \\
    W(x,p_N) &=& \textrm{Tr} \left [\hat{\rho}U(x,N) \right ],
  \end{eqnarray}
 Using the fact that the density matrix is Hermitian, we easily show that the Wigner function is real.
 The Weyl function is in general complex.
 We can easily show the following formulas (which can also be used as alternative definitions):
  \begin{eqnarray}\label{Weyl2}
   \tilde{W}(\alpha,K)&=&\frac{1}{2\pi r} \int_{0}^{2\pi r} \langle x-\frac{\alpha}{2}|\hat{\rho}|x+\frac{\alpha}{2}\rangle
                         \exp \left (i\frac{K}{r}x \right ) \textrm{d}x, \\
                      &=&\sum_{N=-\infty}^{\infty} \langle p_{N-K}|\hat{\rho}|p_N \rangle 
                         \exp \left [-i\alpha \left (p_N-\frac{K}{2r} \right ) \right ].\\
   W(x,p_N) &=& \frac{1}{2\pi r} \int_{0}^{2\pi r} \langle x-\alpha|\hat{\rho}|x+\alpha\rangle
            \exp(i2\alpha p_N) \textrm{d}\alpha,  \\
          &=& \sum_{K=-\infty}^{\infty} \langle p_{N+K}|\hat{\rho}|p_{N-K} \rangle \exp \left (\frac{2ixK}{r} \right ).
  \end{eqnarray}
The Wigner function describes the pseudo-probability of finding the particle in phase space,
in a way consistent with quantum mechanics and the uncertainty principle.
The Weyl function is equal to the overlap of the displaced state with the
original state. In this sense, the $\alpha$, $K$ are position and momentum {\it increments}.
The Weyl function can be understood as a {\it generalised correlation function}.
If we have the wavefunction $R(x)$ 
in order to find the correlation we displace it into $R(x+\alpha)$ and take the integral
of $R(x) R(x+\alpha)$. In the Weyl function we perform a more general displacement in phase space i.e., a displacement 
in both position and momentum. Therefore the correlation is a special case of the Weyl function with
$\alpha=0$ (or $K=0$). We note that the momentum takes values $(N+\sigma)/r$ and depends on $\sigma$.
In contrast the momentum increments appearing on the Weyl function take values $K/r$ and do {\bf not} depend on $\sigma$.

  The Wigner function is related to the Weyl function through a two-dimensional Fourier transform
  \begin{equation}
   W(x,p_N)=\frac{1}{2\pi r} \int_{0}^{2\pi r} \sum_{K=even} \tilde{W}(\alpha,K)
            \exp \left (-i\frac{K}{r}x \right )\exp(i\alpha p_N) \textrm{d}\alpha.
  \end{equation}
 This can be proved using Eq(\ref{Dda}) or Eq(\ref{QW}). This result is intimately connected to the fact that the 
 displacement operator and displaced parity operator are related to each other through a two-dimensional Fourier transform
 (Eq(\ref{QW})).
  
  Using Eqs(\ref{AQpD}),(\ref{PU}) we easily see that
 the Wigner function is periodic function of $x$ with period $\pi r$;
 and the Weyl function is quasi-periodic with `period' $2\pi r$:
 \begin{eqnarray}\label{QW}
   \tilde{W}(\alpha+2\pi rw,K) = (-1)^{Kw} \exp(-i2\pi \sigma w) \tilde{W}(\alpha,K);\;\; \;\;\;\;
   W(x+\pi r,p_N) = W(x,p_N).
 \end{eqnarray}
 The Wigner and Weyl functions depend on the `magnetic flux' $\sigma$ although for simplicity in the
 notation we have not shown this dependence explicitly. They obey the relations
 \begin{eqnarray}\label{Wigsigma}
   \tilde{W}_{\sigma}(\alpha,K) = \tilde{W}_{\sigma +1}(\alpha,K);\;\; \;\;\;\;
   W_{\sigma +1}(x,p_N) = W_{\sigma }(x,p_{N+1}).
 \end{eqnarray}
 This is related to the fact that the momentum is $p_N=\frac{N+\sigma}{r}$
 and as we go from $\sigma$ to $\sigma +1$ the momentum $ p_N$ is relabelled as $ p_{N+1}$.

 The properties of the displacement and parity operator that we proved above can be translated into properties
 for the Wigner and Weyl functions.
 Starting with the Weyl function we use Eqs(\ref{DdK}),(\ref{Dda}) to prove that:
 \begin{eqnarray}
   \sum_{K=-\infty}^{\infty}\tilde{W}(\alpha,K) &=& \langle -\frac{\alpha}{2}|\rho|\frac{\alpha}{2} \rangle; \\
   \frac{1}{2\pi r} \int^{2\pi r}_{0} \tilde{W}(\alpha,K) \exp \left (\frac{i\alpha\sigma}{r} \right )
          \textrm{d}\alpha &=&
          \left\{  \begin{array}{ll}
                  \langle p_M|\rho|p_{-M}\rangle , & K=2M; \\
                   \sum_{N=-\infty}^{\infty} \langle p_N|\rho|p_{N+K}\rangle \Delta \left (-\frac{K}{2}-N \right ), & K=2M+1.
                \end{array}
          \right.  \label{Weylda}
  \end{eqnarray}
 We can make here a similar comment to that made for Eq(\ref{Dda}).
 In Eq(\ref{Weylda}), integration from $2\pi r$ to $4\pi r$, gives the same result for even $K$
and the same result but with opposite sign for odd $K$.
 Therefore if we integrate from $0$ to $4\pi r$
we get the same result for even $K$ and zero for odd $K$. We can also show that
  \begin{eqnarray}
   \tilde{W}(0,0) = 1, \;\;\;\;\;
   \tilde{W}(\alpha,K) = \tilde{W}^*(-\alpha,-K), \;\;\;\;\;
   |\tilde{W}(\alpha,K)| \le 1.
  \end{eqnarray}

 For the Wigner function we use Eqs(\ref{UdK}),(\ref{Uda}),(\ref{UI}) to prove:
  \begin{eqnarray}
   \frac{1}{2\pi r} \int_{0}^{2\pi r} W(x,p_N) \textrm{d}x &=& \langle p_N|\hat{\rho}|p_N \rangle, \\
   \sum_{N=-\infty}^{\infty} W(x,p_N) &=& \frac{1}{2}(\langle x|\hat{\rho}|x \rangle
                                          + \langle x+\pi r|\hat{\rho}|x+\pi r \rangle), \\
   \frac{1}{2\pi r} \int_{0}^{2\pi r} \sum_{N=-\infty}^{\infty} W(x,p_N)\textrm{d}x &=& 1.
  \end{eqnarray}

 We can also prove that
  \begin{eqnarray}
    \textrm{Tr}(\hat{\rho_1}\hat{\rho_2}) &=& \frac{1}{2\pi r} \int_{0}^{2\pi r} \sum_{N=-\infty}^{\infty}
                                       W_{\rho_1}(x,p_N) W_{\rho_2}(x,p_N) \textrm{d}x,
  \end{eqnarray}

\section{Example}

As an example, we consider a `free particle' descibed with the Hamiltonian
\begin{equation}
\hat{H}=\hat{p}^2.
\end{equation}
We note that the particle feels a vector potential through the quasi-periodic boundary condition.

We assume that at $t=0$ the wavefunction is a `Gaussian  on a circle' i.e., a Theta function.
In order to explain this we introduce the
Zak transform \cite{Zak} on a function $S(y)$ on a real line, defined as
\begin{equation}
R(x,\sigma)= {\cal N} \sum_{w=-\infty}^{\infty} S(y=x+2\pi rw)\exp (-i2\pi \sigma w), \label{qperiodic}
\end{equation}
where ${\cal N}$ is used to normalize the function $R(x,\sigma)$ according to Eq(\ref{Norm}).
If $S(y;A)$ is a Gaussian wavefunction
\begin{equation}\label{Gauss}
S(y;A) = \pi^{-1/4}\exp \left (-\frac{y^2}{2}+\sqrt{2}A\cdot y-A A_R\right )
\end{equation}
where $A=A_R+iA_I$, then the Zak transform is
\begin{equation}\label{Gaucir}
R(x,\sigma;A)={\cal N} \cdot \pi^{-1/4}\exp \left (-\frac{x^2}{2}+\sqrt{2}Ax-A A_R\right )
\Theta_3\left[-\pi \sigma+i\pi r(x-\sqrt{2}A);i2\pi r^2\right ]
\end{equation}
where $\Theta_3[u;\tau]$ is Theta function \cite{theta}, defined as
\begin{equation}
\Theta_3[u;\tau]=\sum_{n=-\infty}^{\infty}\exp (i\pi \tau n^2+i2nu).
\end{equation}
We have taken $r=1$, $A=1$ and calculated the Wigner
and Weyl functions. Numerical results are shown in Figs 1-3. 

In Fig.1 we present a contour plot of the Wigner function $W(x,N+\sigma)$ as a
function of $x$ and for {\bf all} values of $\sigma$.
For a given $\sigma$ the Wigner function is defined only on the discrete values of the momenta
$p_N=(N+\sigma)/r$ and the parallel black lines in the figure show the case $\sigma=0.1$.
As expected from Eq(\ref{QW}), the Wigner function is
periodic function of $x$ with period $\pi $. Also it has been
explained in Eq(\ref{Wigsigma}) that the plotted values of
$W(x,p_1)$ for $\sigma=1$ represent the $W(x,p_2)$ for $\sigma
=0$. 

In Fig.2 we present the Wigner function as a function of $x$ at 
a particular momentum $p_0=\sigma$ with $\sigma=0.1$. This is really an appropriate slice 
of the three-dimensional version of Fig.1.

In Fig.3 we present a contour plot of the absolute value of the Weyl function
as a function of $\alpha$ for $K=1$ and for {\bf all} values of $\sigma$.
The black line in the figure shows the case $\sigma=0.1$.

We calculate the time evolution of this system. This is easily done in the momentum representation
as
\begin{eqnarray}\label{A}
\langle p_N|R(t) \rangle &=& \exp(it\hat{H}) \langle p_N|R(0)\rangle   
\end{eqnarray}
We need the state of the system at $t=0$ in the momentum representation.
If $\tilde{S}(p)$ is the Fourier transform of $S(x)$:
   \begin{equation}
    \tilde{S}(p)=\int^{+\infty}_{-\infty} S(x) \exp(-ipx) \textrm{d}x, \label{FT}
   \end{equation}
then using Eq(\ref{Rn}), Eq(\ref{qperiodic}) and Eq(\ref{FT}) we can prove the relation:
   \begin{equation}
    R_N=\frac{{\cal N}}{2\pi r}\tilde{S}(p_N).  \label{RSp}
   \end{equation}
Since the Fourier transform of Gaussian wavefunction Eq(\ref{Gauss}) is:
  \begin{equation}
   \tilde{S}(p;A)= \sqrt{2}\pi^{1/4} \exp \left (-\frac{p^2}{2}-i\sqrt{2}A\cdot p + A A_I\right ),
  \end{equation}
we obtain:
  \begin{equation}\label{B}
    \langle p_N|R \rangle =\frac{{\cal N}\pi^{1/4}}{\sqrt{2}\pi r} \exp \left (-\frac{{p_N}^2}{2}-i\sqrt{2}A\cdot p_N + A A_I\right ).  
  \end{equation}
Inserting Eq(\ref{B}) into (\ref{A}) we get
  \begin{eqnarray}
    \langle p_N|R(t) \rangle 
      = \frac{{\cal N}\pi^{1/4}}{\sqrt{2}\pi r} \exp \left [-\left ( \frac{1}{2}-it\right ){p_N}^2 -i\sqrt{2}A\cdot p_N + A A_I\right ].
  \end{eqnarray}
From that we have calculated for $t=1$ the corresponding Wigner and Weyl functions to the previous ones.
Results are shown in Figs 4-6.
The results show the time evolution of the system in the language of Wigner and weyl functions.

\section{Discussion}
Quantum mechanics on a circle has attracted a lot of attention in the literature. It uses topological ideas in the 
context of quantum mechanics. In this paper we have studied phase space methods in this context. We have introduced
displacement operators and studied their properties. There is clearly some analogy with the harmonic oscillator 
displacement operator properties (given in Eqs(\ref{HD1}),(\ref{HD2}),(\ref{HD3})), but there are also considerable differences. 
In particular the flux $\sigma$ plays an important role in the properties of the displacement operators for particles on a circle.
 Our main results here are Eqs(\ref{DdK}),(\ref{Dda}),(\ref{DdaKev}).

We also introduced Wigner and Weyl funcitons and discussed their physical interpreration and their properties, which are 
direct consequence of the displacement operator properties. A numerical example for the Theta wavefunction of Eq(\ref{Gaucir}) 
(the analogue on a circle of the Gaussian function on the real line) has been discussed.

The results can be used in the context of Aharonov-Bohm devices, Floquet-Bloch solid state systems, 
coherent states on a circle \cite{Comm}, quantum maps and classical and quantum chaos \cite{chaos} etc. Fractional Fourier
transforms in this context have been discussed in \cite{chong}.

\begin{figure}[p]
    \begin{center}
     \includegraphics[scale=0.8]{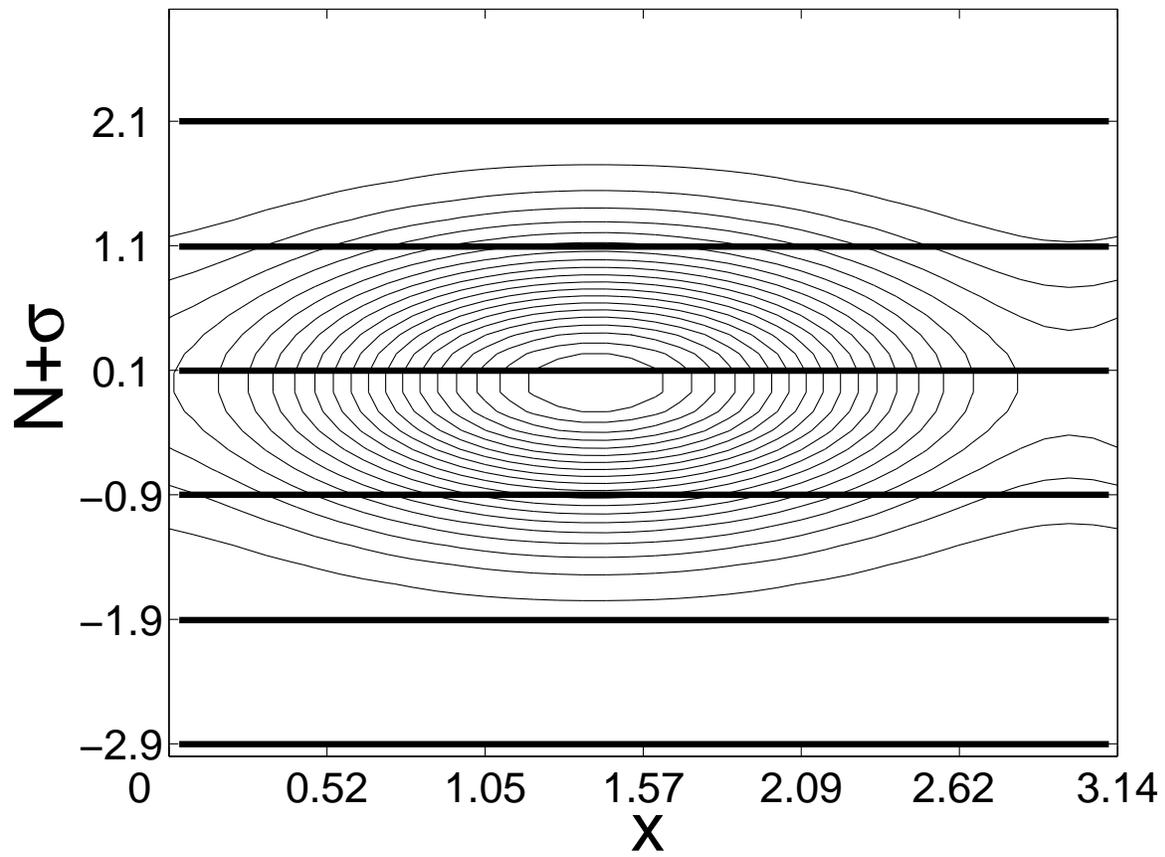}
     \caption{A contour plot of the Wigner function at $t=0$ as a function  
     of $x$ and for all values of $\sigma$.
The parallel black lines in the figure show the case $\sigma=0.1$.}
     \label{Fig1}
\end{center}
\end{figure}

\begin{figure}[p]
    \begin{center}     
     \includegraphics[scale=0.8]{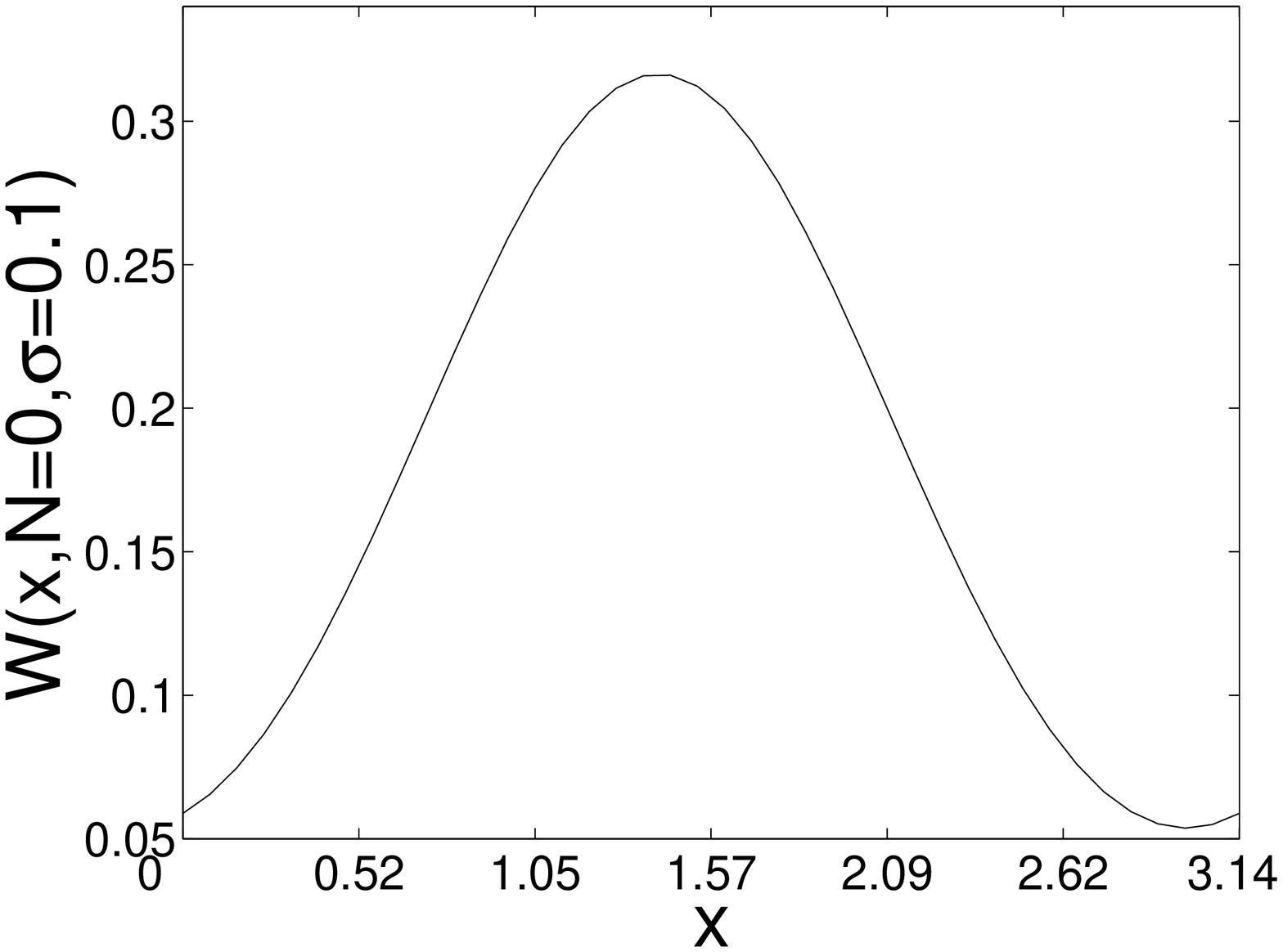}
     \caption{A slice of the Wigner function of Fig.1 with $\sigma=0.1$}
     \label{Fig2}
    \end{center}
\end{figure}

\begin{figure}[p]
    \begin{center}
     \includegraphics[scale=0.8]{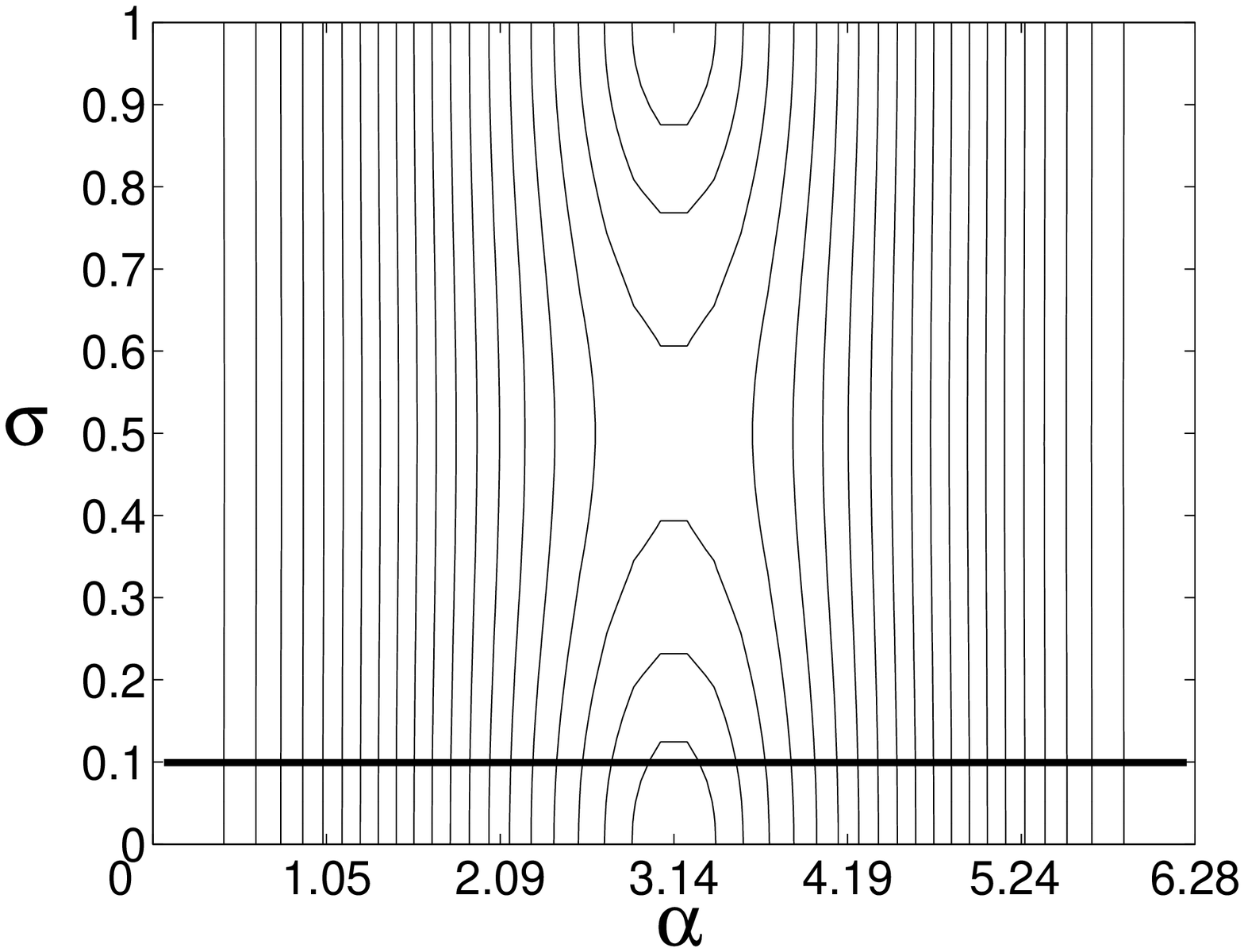}
     \caption{A contour plot of the absolute value of the Weyl function at $t=0$
     for $K=1$, as a function of $\alpha$
               and $\sigma$. The black line in the figure shows the case $\sigma=0.1$ }
     \label{Fig3}
    \end{center}
\end{figure}

\begin{figure}[p]
    \begin{center}
     \includegraphics[scale=0.8]{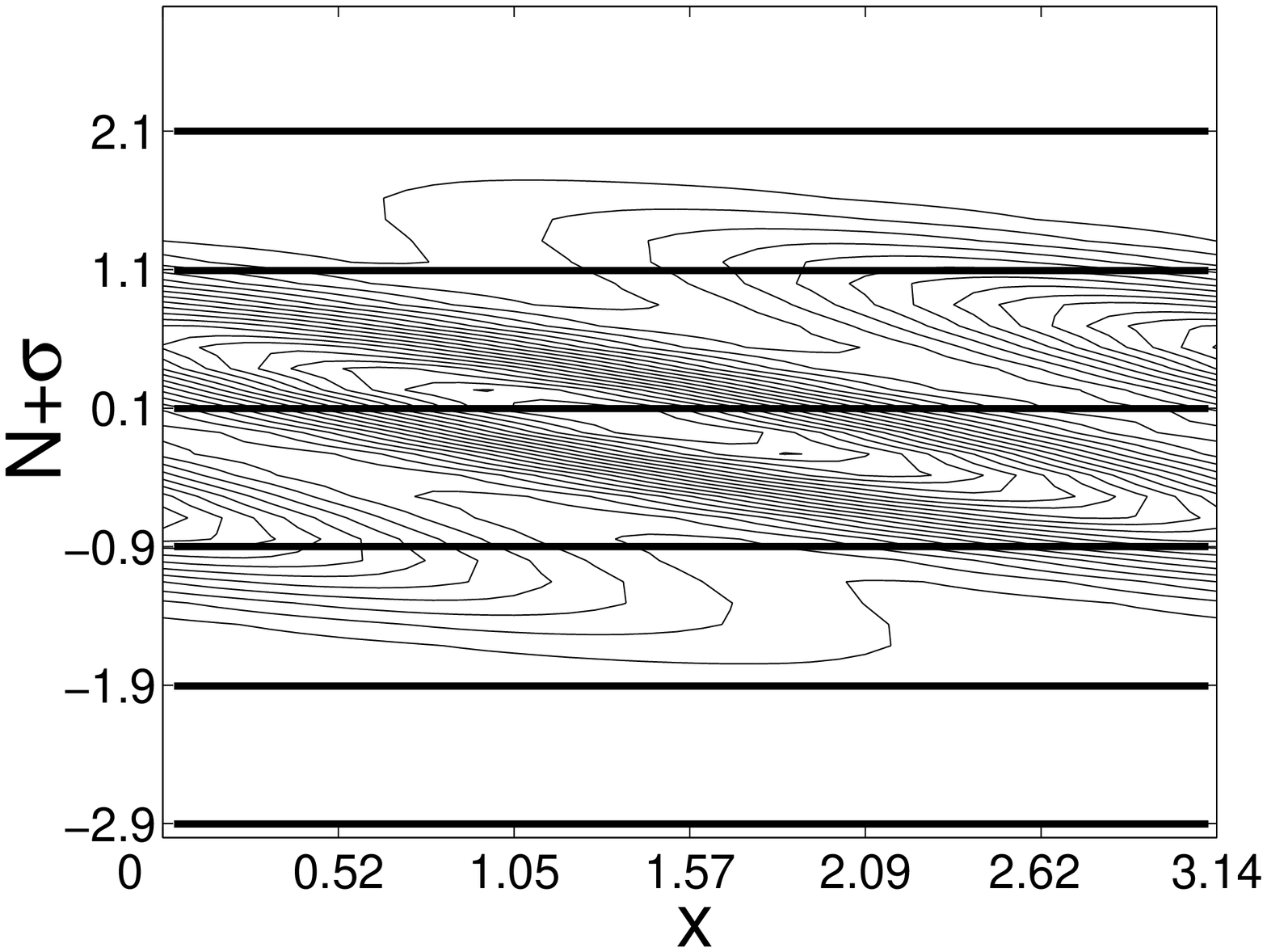}
     \caption{A contour plot of the Wigner function at $t=1$ as a function  
     of $x$ and for all values of $\sigma$.
The parallel black lines in the figure show the case $\sigma=0.1$}
  
     \label{Fig4}
     
 \end{center}
\end{figure}

\begin{figure}[p]
    \begin{center} 
     \includegraphics[scale=0.8]{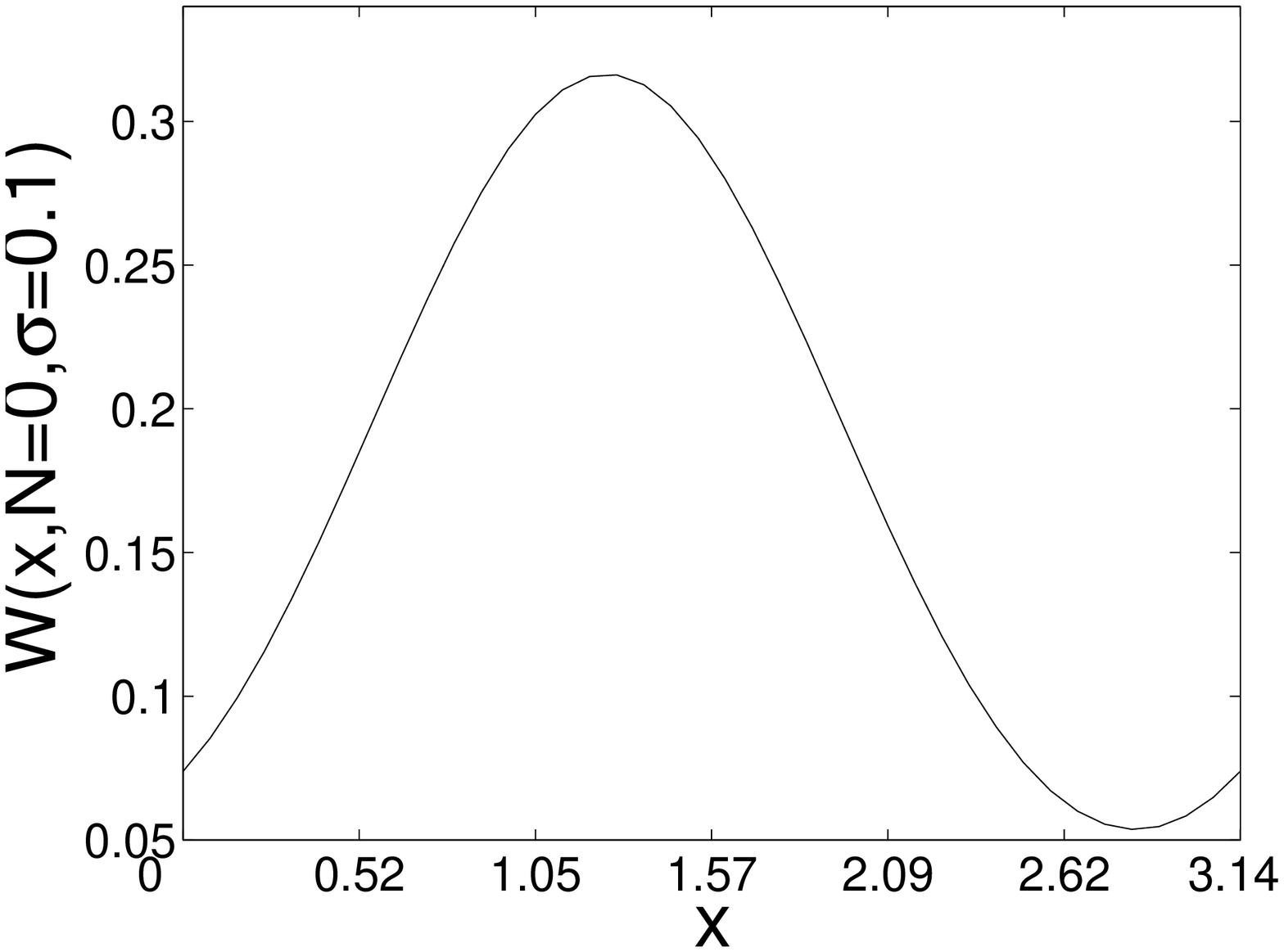}
     \caption{A slice of the Wigner function of Fig.4 with $\sigma=0.1$}
     
     \label{Fig5}
    \end{center}
\end{figure}

\begin{figure}[p]
    \begin{center}
     \includegraphics[scale=0.8]{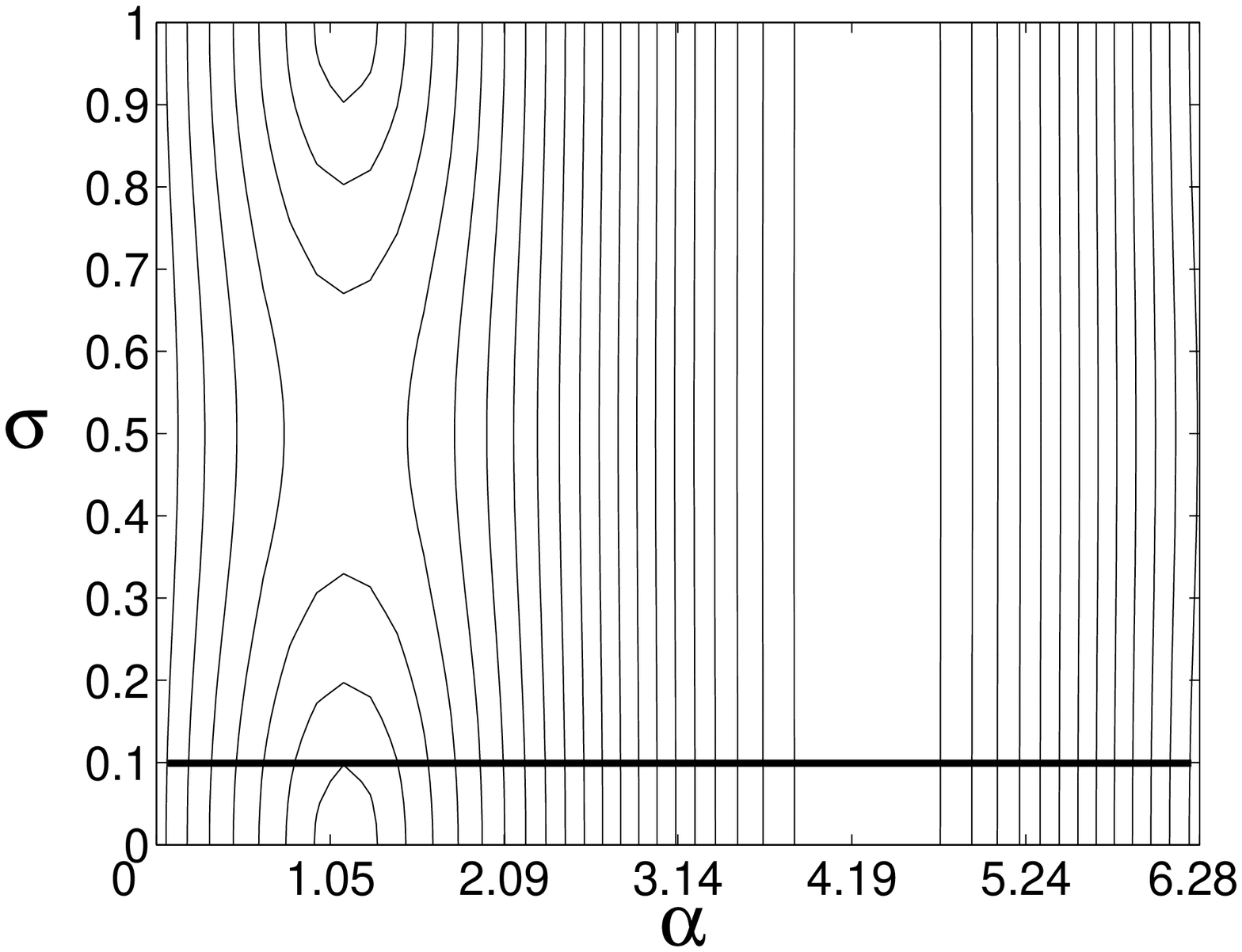}
     \caption{A contour plot of the absolute value of the Weyl function at $t=1$
     for $K=1$, as a function of $\alpha$
               and $\sigma$. The black line in the figure shows the case $\sigma=0.1$}
     \label{Fig6}
    \end{center}
\end{figure}

\end{document}